\documentstyle[12pt,aaspp4]{article}
\begin{document}

\title{TWO-TEMPERATURE INTRACLUSTER MEDIUM IN MERGING CLUSTERS OF GALAXIES}
\author{{\sc Motokazu Takizawa}}
\affil{Department of Astronomy, Faculty of Science, Kyoto University, 
       Sakyo-ku, Kyoto 606-8502, JAPAN}
\authoremail{takizawa@kusastro.kyoto-u.ac.jp}

\begin{abstract}
We investigate the evolution of intracluster medium during 
a cluster merger, explicitly considering the relaxation 
process between the ions and electrons by N-body and hydrodynamical 
simulations. When two subclusters collide
each other, a bow shock is formed between the centers of 
two substructures and propagate in both directions along the collision axis.
The shock primarily heats
the ions because the kinetic energy of an ion entering the shock is 
larger than that of an electron by the ratio of masses. In the 
post-shock region the energy is transported from the ions to electrons
via Coulomb coupling.
However, since the energy exchange timescale depends both on the gas density 
and temperature, distribution of electron temperature becomes 
more complex than that of the plasma mean temperature,
especially in the expanding phase.
After the collision of two subclusters, gas outflow occurs not only
along the collision axis but also in its perpendicular direction. 
The gas which is originally located in the central part of the subclusters 
moves both in the parallel and perpendicular directions. 
Since the equilibrium timescale of 
the gas along these directions is relatively short, temperature difference 
between ions and electrons is larger in the directions tilted
by the angles of $\pm 45^\circ$ with respect to the collision axis.
The electron temperature could be significantly lower that the plasma mean
temperature by $\sim 50 \%$ at most. The significance of our results in the
interpretation of X-ray observations is briefly discussed.
\end{abstract}


\section{INTRODUCTION}
Clusters of galaxies (CG) contain collisionless particles, galaxies and dark
matter, and a diffuse gas component. The gas component is called
intracluster medium (ICM). The ICM is the plasma with temperature of
about $10^8$ K, thus emitting X-ray mainly through the thermal
bremsstrahlung of the electrons (Sarazin 1988). Since the ICM is optically
thin for X-rays and spreads over the entire cluster, it can provide good 
information regarding the physical property of the clusters.

According to the hierarchical clustering scenario, it is believed
that CG are formed through subcluster mergers and/or absorption of 
smaller galaxy groups. The rate of mergers are dependent on 
the cosmological parameters, especially on the density parameter,
$\Omega_0$ (Richstone, Loeb, \& Turner 1992). 
However, when this is applied to more
realistic situations such as observational results
(eg., West \& Bothun 1990; Rhee et al. 1991; Jones \& Forman 1992; 
Mohr, Fabricant, \& Geller 1993; Bird 1994; 
Serna \& Gerbal 1996;  Gurzadyan \& Mazure 1998; 
Solanes, Salvador-Sol\'{o}, \& Gonz\'alez-Casado 1998)
or cosmological numerical simulations
(Evrard et al. 1993; Jing et al. 1995; Crone, Evrard, \& Richstone, 1996), 
uncertainty of dynamical property of CG 
during mergers seriously affects the estimation of $\Omega_0$ 
(Nakamura, Hattori, \& Mineshige 1995).

To study evolution of CG, plenty of numerical simulations have been 
performed so far using N-body and hydrodynamical codes. Perrenod (1978)
was the first to calculate the evolution of a spherical symmetric CG with a 
standard mesh hydrodynamical code. Recently, Takizawa\& Mineshige (1998)
calculated the spherical CG and investigate the effect of pre-heating 
on ICM density profiles. To investigate the formation and evolution 
in more realistic situations, many simulations starting from the 
cosmological initial density perturbations are performed (e.g., 
Evrard 1990; Katz \& White 1993; Metzler \& Evrard 1994; 
Navarro, Frenk, \& White 1995; Bryan \& Norman 1998; 
Eke, Navarro, \& Frenk 1998; Yoshikawa, Itoh, \& Suto 1998).
Schindler \& M\"{u}ller (1993) find that characteristic temperature 
structures occur in merging clusters through shock heating and 
adiabatic compression and expansion.
On the other hand, there are simulations starting from 
rather idealized initial conditions to study the mergers in detail.
Roettiger, Loken, \& Burns (1997) study general behaviors of 
head on collisions with a high resolution finite-difference code.
Their observational consequences are discussed in more detail in
Roettiger, Burns, \& Loken (1996). Ishizaka (1997) find that
the specific energy ratio of ICM and galaxies, $\beta_{\rm spec}$, is a
good indicator of merging clusters and that it can be used to 
determine the phase of the merger. In particular, off-center collisions
are investigated by Ricker (1998).

In recent years, rather complex electron temperature structures are 
revealed in merging CG through X-ray observations (e.g., 
Fujita et al. 1996; Honda et al. 1996; Churazov et al. 1998; 
Donnelly et al. 1998a; Donnelly et al. 1998b; Davis \& White 1998;
Markevitch, Sarazin, \& Vikhlinin 1998).
In addition, Hanami et al. (1998) find that the energy of iron K $\alpha$ 
line is higher than that expected from the temperature in ionization 
equilibrium plasma in SC1329-313 in Shapley concentration. These
results implicate there exits strong bulk flow motion in ICM 
of merging clusters.
Temperature maps of ICM contain important information of mergers
as well as $\beta_{\rm spec}$ does. Some numerical simulations
are tried to explain these observational results; Roettiger, Burns, \& 
Pinkney (1995) tried the modeling for A2256, and Ishizaka \& Mineshige 
(1996) for Coma. More detailed ``anatomy'' is  done for A754 by
Roettiger, Stone, \& Mushotzky (1998).

As mentioned above, temperature structures of ICM provide us with 
very important
information about the phase of the mergers. However, we can find 
only the electron temperature through X-ray observations until now. 
In all the previous studies the discussion is based on the assumption 
that observed electron temperature ($T_{\rm e}$) is the same as 
the plasma mean temperature ($\bar{T}$).
However, we believe this assumption is problematic.
The shock primarily heats the ions because the kinetic
energy of an ion entering the shock is larger than that of an electron 
by the ratio of masses. In addition, the equilibrium timescale between 
electrons and ions ($t_{\rm eq}$) in typical CG becomes $10^8 \sim 10^9$ yr
if Coulomb coupling is considered as the relaxation process
(see Spitzer 1962):
\begin{eqnarray}
  t_{\rm eq} = 2.0 \times 10^8 {\rm yr} 
           \biggr( \frac{\ln \Lambda}{40} \biggl) ^{-1}
           \biggr( \frac{n_{\rm i}}{10^{-3} {\rm cm}^{-3}} \biggl)^{-1}
           \biggr( \frac{T_{\rm e}}{10^8 {\rm K}} \biggl)^{3/2}, 
	   \label{eq:teq}
\end{eqnarray}
where $n_{\rm i}$ is the ion density, $T_{\rm e}$ is the electron
temperature, and $\ln \Lambda$
is the Coulomb logarithm. Indeed, some CG have larger electron temperature
gradients than those expected in numerical simulations (Markevitch 1996).
A two-temperature model of ICM is one of solutions to solve this discrepancy
(Fox \& Loeb 1997; Chi\'eze, Alimi, \& Teyssier, 1998; 
Ettori\&Fabian 1998; Takizawa 1998). 
In cluster mergers, it is believed that shock heating 
plays an important role in the evolution of ICM. Furthermore, according 
to the previous studies about mergers, timescale surviving complex 
temperature structures is an order of $10^9$ yr, which is comparable to 
$t_{\rm eq}$. Therefore, difference between $T_{\rm e}$ and $\bar{T}$ 
should be considered properly to study the evolution of $T_{\rm e}$ 
structure in mergers.

To investigate the evolution of two-temperature ICM in cluster mergers,
we perform N-body and hydrodynamical simulations considering the relaxation 
process between ions and electrons. For the hydrodynamical part, we choose
the smoothed particle hydrodynamics (SPH) method. Although SPH codes are
not better for problems with shocks than mesh codes, its Lagrangean nature
is suitable to deal the electron-ion coupling simply. 

The rest of this paper is organized as follows. In \S \ref{s:ae} we estimate
analytically some physical quantities relevant to cluster mergers. 
In \S \ref{s:sim} we describe the adopted numerical methods and initial
conditions for our simulations. In \S \ref{s:res} we present the results.
In \S \ref{s:sd} we summarize the results and discuss their implications.

\section{BASIC CONSIDERATIONS OF EQUILIBRIUM TIMES}\label{s:ae}
In this section, we consider the scaling laws of CG and the 
initial conditions of cluster mergers.
We assume that the power spectrum of the cosmological density fluctuation
field, $P(k)$, obeys
power law, $P(k) \propto k^n$. According to the spherical model 
for the non-linear collapse (see Peebles 1980, Padmanabhan 1993, etc.), 
the virial radius ($r$), the density ($\rho$), and the virial temperature
($T$) of the dark halo obey the scaling laws as follows,
\begin{eqnarray}
	r \propto M^{(n+5)/6},  \\
	\rho \propto M^{-(n+3)/2}, \\
	T \propto M^{(1-n)/6},
\end{eqnarray}  
where $M$ is the total mass of the halo. The effective index of $P(k)$
of CDM is $-1 \sim -2$ in the galactic and CG scale. We consider
the case of $n=-2$. Then the above relations are,
\begin{eqnarray}
	r \propto M^{1/2}, \label{eq:screlofr} \\
	\rho \propto M^{-1/2}, \\
	T \propto M^{1/2}.
\end{eqnarray}  
Using these relations and the gas fraction, $f_{\rm g}$, 
we obtain the scaling laws 
of equilibrium timescale between ions and electrons, 
$t_{\rm eq} \propto (f_{\rm g} \rho)^{-1} T^{3/2}$, as follows.
\begin{eqnarray}
	t_{\rm eq} \propto f_{\rm g}^{-1} M^{5/4}.
        \label{eq:screlteq}
\end{eqnarray}  
Thus, the more massive CG is, the longer $t_{\rm eq}$ is,
if $f_{\rm g}$ is constant.
However, the observational results about the luminosity-temperature (LT) 
relation suggests $f_{\rm g}$ depends on $M$ 
(Mushotzky 1984; Edge \& Stewart 1991;
David et al. 1993; Fukazawa 1997; Mushotzky \& Scharf 1997; Markevitch 1998; 
Arnaud \& Evrard 1998). Since $L \propto (f_{\rm g} \rho)^2 T^{1/2} R^3$, 
$f_{\rm g} \propto M^{3/8}$ if we assume $L \propto T^3$. Then we obtain
\begin{eqnarray}
	t_{\rm eq} \propto M^{7/8}.
\end{eqnarray}  
$t_{\rm eq}$ is longer in more massive CG also in this case,
though the dependence is slightly weaker than the case when $f_g$ is constant.

Next, let us consider the the merger of subclusters whose masses are
$M_1$ and $M_2$, respectively. The collision velocity $V$ is 
estimated as follows.
From the dynamical energy conservation law, we obtain,
\begin{eqnarray}
	- \frac{G M_1 M_2}{2 R} = - \frac{G M_1 M_2}{r_1+r_2} + 
        \frac{1}{2} M V^2  \label{eq:dyenconl},
\end{eqnarray}  
where $G$ is the gravitational constant, $r_1$ and $r_2$ is the virial radius
of the each subcluster, $R$ is the maximum expansion radius of the 
whole system, and $M=M_1 M_2/(M_1+M_2)$ is the reduced mass.
Since the maximum expansion radius is twice the virial radius, using 
the relation (\ref{eq:screlofr}) we obtain
\begin{eqnarray}
	R = 2 \times \biggr( \frac{M_1 + M_2}{M_1} \biggl)^{1/2} r_1. 
        \label{eq:larger}
\end{eqnarray}  
It is convenient to introduce the parameter $\alpha \equiv M_2/M_1$.
Without losing the generality, we can set $0<\alpha<1$. Then, from 
the scaling relations of (\ref{eq:screlofr}) and equations 
(\ref{eq:dyenconl}) and (\ref{eq:larger}), we can describe $V$ as follows,
\begin{eqnarray}
	V^2 = \frac{2GM_1}{r_1} (1+\alpha) \biggr[ \frac{1}{1+\alpha^{1/2}}
              - \frac{1}{4(1+\alpha)^{1/2}} \biggl].
\end{eqnarray}  
On the other hand, the sound speed of subcluster 1, $c_1$ is
\begin{eqnarray}
	c_1^2 = \gamma (\gamma-1) \frac{G M_1}{2 r_1}, 
\end{eqnarray}  
where $\gamma$ is the ratio of specific heat. Thus, the Mach number,
${\cal M}_1=V/c_1$,  is,
\begin{eqnarray}
	{\cal M}_1^2 = \frac{4(1+\alpha)}{\gamma(\gamma-1)} \biggr[
            \frac{1}{1+\alpha^{1/2}}-\frac{1}{4(1+\alpha)^{1/2}} \biggl].
\end{eqnarray}  
From the scaling relation (\ref{eq:screlofr}), 
the Mach number for subcluster 2, ${\cal M}_2$, is,
\begin{eqnarray}
       {\cal M}_2^2 = \alpha^{-1/2} {\cal M}_1^2.
\end{eqnarray}
Figure \ref{fig:mach} shows ${\cal M}_1^2$ (solid line) and 
${\cal M}_2^2$ (dotted line) with 
respect to $\alpha$ when $\gamma=5/3$. For major mergers ($\alpha>0.1$), 
these values are greater than $\sim 2$. Thus, in CDM universe, supersonic 
collisions are quite natural in major mergers. 

The Mach number gives the maximum temperature difference between 
ions and electrons in the post-shock regions. 
The specific inertial energy of ions and electrons are,
in the limit of no mutual energy exchange,
\begin{eqnarray}
	\frac{k_{\rm B} T_{\rm i}}{\mu m_{\rm p}} &\sim& c^2+V^2, \\
	\frac{k_{\rm B} T_{\rm e}}{\mu m_{\rm p}} &\sim& c^2,
\end{eqnarray}
where $\mu$ is the mean molecular weight, 
$m_{\rm p}$ is the proton mass, $k_{\rm B}$ is the Boltzmann constant,
and $c$ is the sound speed in the pre-shock regions. Thus the difference
between $T_{\rm e}$ and $\bar{T} \sim (T_{\rm i}+T_{\rm e})/2$, is,
\begin{eqnarray}
	\frac{T_{\rm e}}{\bar{T}} \sim \frac{1}{1+{\cal M}^2/2}. \\
\end{eqnarray}
Thus, from figure \ref{fig:mach}, we expect $T_{\rm e}/\bar{T}$ is less than 
about $0.6$. We will confirm this later by numerical simulations.

\section{THE SIMULATIONS}\label{s:sim}

In the present study, we consider the CG consisting of two components:
collisionless particles corresponding to galaxies and DM and 
two-temperature gas 
corresponding to the ICM. When calculating gravity, both components are
considered, although the former dominates over the latter.
Radiative cooling and heat conduction are not included in our 
simulations.

\subsection{The Numerical Method}
To solve the hydrodynamical equations for the gas component, we used the 
smoothed-particle hydrodynamics (SPH) method (see Monaghan 1992). 
Although more accurate finite-difference methods better handle shocks 
than SPH methods, its fully Lagrangean nature is convenient to calculate the 
evolution of two-temperature plasmas since the coupling between ions 
and electrons can be treated more simply in Lagrangean view as follows.

As the standard SPH codes for one-temperature fluid, we solve
the continuity equation, the momentum equation, and the thermal energy
equation with artificial viscosity to treat the shocks. 
In addition to these equations, we solve one more equation for the 
normalized electron temperature, 
$\tilde{T}_{\rm e} \equiv T_{\rm e}/\bar{T}$, 
where $T_{\rm e}$ is 
the electron temperature and $\bar{T}$ is the plasma mean temperature. 
We assume that artificial viscous heating is effective only for ions and that
only the Coulomb coupling is considered in the relaxation process. Then,
the Lagrangean time evolution of $\tilde{T}_{\rm e}$ is (see Appendix),
\begin{eqnarray}
	\frac{ d \tilde{T}_{\rm e} }{dt} = 
             \frac{ \tilde{T}_{\rm i} - \tilde{T}_{\rm e} } 
             {t_{\rm eq}} - \tilde{T}_{\rm e} \frac{Q_{\rm vis}}{u}, 
	\label{eq:techevo}
\end{eqnarray}
where $\tilde{T}_{\rm i} \equiv T_{\rm i}/\bar{T}$ is the normalized 
ion temperature, $Q_{\rm vis}$ is the artificial viscous heating 
per unit mass, and $u$ is the thermal energy per unit mass. 
For the numerical integration of equation (\ref{eq:techevo}), we should
take a time-step shorter enough than not only the Courant and viscous timescale
as usual SPH codes (see Monaghan 1992 \S10.3) but also $t_{\rm eq}$. 
In our simulations, however, the latter is much shorter than the former 
in the central high-density regions. Thus, the integration with timestep
relevant to $t_{\rm eq}$ takes a huge computational time. In addition, 
since two-temperature nature is not important in such regions, 
the time-step control above mentioned is not a good choice.
To integrate equation (\ref{eq:techevo}) with the Courant and the viscous
time-step control, we use the results of Fox \& Loeb (1997) as follows.
Let us integrate equation (\ref{eq:techevo}) from $t=t_0$ to $t=t_0+\Delta t$.
First, we integrate equation (\ref{eq:techevo}) regarding the second term 
on the right-hand side being negligibly small. 
In this case, neglecting the small change of Coulomb logarithm 
in $t_{\rm eq}$, we can integrate this analytically
(Fox \& Loeb 1997),
\begin{eqnarray}
	\frac{\Delta t}{t_{\rm 2s}} = \frac{n_{\rm i}}{n_{\rm i}+n_{\rm e}} 
        \biggr[
         F( \tilde{T}_{\rm e} (t_0+\Delta t) ) - F( \tilde{T}_{\rm e}(t_0) )
        \biggl],
\end{eqnarray}
where, 
\begin{eqnarray}
	F(x) = \ln \biggr( \frac{1+\sqrt{x}}{1-\sqrt{x}} \biggl)
              - 2 \sqrt{x} \biggr( 1 + \frac{x}{3} \biggl),
\end{eqnarray}
and $t_{\rm 2s} \equiv t_{\rm eq}(t_0) \tilde{T}_e(t_0)^{-3/2}$.
Second, we add the contribution from the second term to 
$\tilde{T}_{\rm e} (t_0+\Delta t)$ using the second-order Runge-Kutta method.
If the time step is chosen correctly, the displacement of the position of 
each SPH particle is smaller enough than the spatial resolution and 
$Q_{\rm vis}/u$ is smaller enough than unity. 
Thus the above method can follow the evolution of $\tilde{T}_{\rm e}$
efficiently and with reasonable accuracy.

Gravitational forces are calculated by the Barnes-Hut tree algorithm
(Barnes \& Hut 1986) and softened using a Plummer potential profile
$\Phi \propto (r^2+\epsilon^2)^{-1}$, where $\epsilon$ is the softening
parameter. We set $\epsilon$ one tenth of the initial core radius 
of the smaller
subcluster in the simulation. Tree structure is also used to search 
for nearest neighbors in SPH calculations (Hernquist \& Katz 1989).
In the standard one-dimensional shock tube test, our code nicely reproduced
the analytic solution. The shock and contact discontinuity is broadened
over a range of about three times of smoothing length.
During the calculations described below, the total energy is conserved within
$\sim 0.4\%/$($1000$  steps).

\subsection{Models and Initial Conditions}

We consider mergers of two virialized subclusters of galaxies with masses
$M_1$ and $M_2$. Initial configuration of each subcluster is described 
as follows. The spatial distribution of the DM in each subcluster is
represented by the King distribution;
\begin{eqnarray}
    \rho_{\rm DM} (r) = \left\{
                          \begin{array}{@{\,}ll}
        \rho_0[1+(r/r_{\rm c})^2]^{-3/2}  &  \mbox{ ($r \le r_{\rm out}$) } \\
                               0     &  \mbox{ ($r > r_{\rm out}$) }           
            		  \end{array}
                    \right. 
\end{eqnarray}
where, $\rho_0$ is the central density, $r_{\rm c}$ is the core radius, and
$r_{\rm out}$ is the radius of the cluster. We set 
$r_{\rm out} = 5 \times r_{\rm c}$. 
This is nearly equal to the virial radius of the King distribution.
$\rho_0$ is determined by the total mass of DM
($M_{\rm DM}$),
\begin{eqnarray}
	\rho_0 = \frac{M_{\rm DM}}{4 \pi r_{\rm c}^3} \biggr[
                 -\frac{t_{\rm out}}{\sqrt{t_{\rm out}^2 + 1}}
                 + \ln \bigr| t_{\rm out}+\sqrt{t_{\rm out}^2 + 1} \bigl|
                                            \biggl]^{-1},
\end{eqnarray}
where $t_{\rm out} = r_{\rm out}/r_{\rm c}$. 
We assume velocity distribution of the DM 
particles to be an isotropic Maxwellian. Then, from the second 
moment of the collisionless Boltzmann equation, one-dimensional velocity 
dispersion at $r$, $\sigma_1^2(r)$, is (e.g., Binney \& Tremaine 1987)
\begin{eqnarray}
	\sigma_1^2 (r) = \frac{1}{\rho_{\rm DM}(r)} 
        \int_r^{r_{\rm out}} \rho_{\rm DM} (r) \frac{G M_r}{r^2} dr,
\end{eqnarray}
where $M_r$ is the mass inside $r$.
We assume that initial ICM temperature is isothermal and equal to 
the virial temperature at $r_{\rm out}$.
The ICM is initially in hydrostatic equilibrium within a cluster potential
of the DM and the ICM, itself.

We set the initial conditions as follows. Two subclusters are placed 
at rest with a separation between the center of each subcluster, $r_{\rm sep}$,
initially. The coordinate system is taken in such a way that
the center of masses is at rest in the origin.
Two representative cases are examined in the preset study:
a collision between two equal-mass clusters (Run A), an absorption of 
a smaller cluster by a larger one (Run B). The scaling law between 
the two subclusters and $r_{\rm sep}$ are determined as the case of 
$P(k) \propto k^{-2}$ in \S\ref{s:ae}. 
In Run A, two subcluster have the same masses; 
$M_1 = M_2 = 0.5 \times 10^{15} M_{\odot}$. We set the core radii to be
$r_{\rm c} = 0.2$Mpc. Each subcluster consists of 5000 collisionless particles
and 5000 SPH particles. The total gas mass fraction is 10\% for both 
clusters. In Run B, the mass ratio is $M_1 : M_2 = 4:1$. The larger
cluster has the same mass as that in Run A. 
The particle numbers of the smaller one are one fourth of the larger one.
The parameters in our calculations are summarized in table \ref{tab:para}.

\section{RESULTS}\label{s:res}

\subsection{Time evolution of the total feature during the mergers}
\label{ss:ltb}

Figure \ref{fig:timea} shows the time evolution of various physical 
quantities representing the total feature of the system of Run A: 
from the top, the total luminosity ($L$), 
the emissivity-weighted mean electron temperature ($T_{\rm e,ew}$), 
the square root of the velocity 
dispersion parallel to the collision axis ($\sigma_{//}$), and the specific
energy ratio of DM and ICM ($\beta_{\rm spec}$). 
$\beta_{\rm spec}$ is defined as,
\begin{eqnarray}
	\beta_{\rm spec} \equiv \frac{\mu  m_{\rm p} \sigma^2_{//} }
                                     {k_{\rm B} T_{\rm e,ew}}.
\end{eqnarray}
When calculating $L$, we assume the thermal bremsstrahlung of 
optically thin plasma (Rybicki \& Lightman 1979) and neglect 
the line emissions.

Around the collision, all of $L$, $T_{\rm e,ew}$ and $\sigma_{//}$ 
rise but on different timescales. Since the rise
of $T_{\rm e,ew}$ is quicker than that of $\sigma_{//}$, $\beta_{\rm spec}$
has two peaks in the time evolution. The minimum of $\beta_{\rm spec}$ is
coincident with the maximum of $T_{\rm e,ew}$. 

Similar results are seen also in Run B (figure \ref{fig:timeb}).
However, the amplitudes of time fluctuation are smaller than 
in Run A. 

Since the emissivity is proportional to the square of the gas density, 
emissivity-weighted mean temperature (EWMT) is more like the temperature 
in the central high density regions, where $t_{\rm eq}$ is $\sim 10^8$yr or
less.  Thus, the difference in EWMTs between
electrons and ions is practically 
negligible. Indeed, this difference becomes
$\sim 1 \%$ even at maximum in both runs. Therefore, our results 
agree qualitatively with those of the previous one-temperature simulations
such as Ishizaka (1997).

\subsection{Time evolution of temperature distribution of Run A}

As seen in \S \ref{ss:ltb}, two-temperature nature of ICM does not 
manifest itself in EWMT of CG. This is not the case, however, 
when we can resolve the spatial distribution of ICM temperature.
Let us examine first the case of Run A.

First of all, let us see the $\bar{T}$ map and the gas velocity field, and 
discuss the dynamical evolution of Run A.
Figure \ref{fig:tbarmapa} shows the snapshots of X-ray surface brightness
(contours) and emissivity-weighted $\bar{T}$ (colors) distribution seen 
from the direction perpendicular to the collision axis.
X-ray surface brightness contours are equally spaced on a logarithmic scale
and separated by a factor of 7.4. The blue, green, yellow, and red colors
correspond to $k_{\rm B}T \sim 5$ keV, $10$ keV, $15$ keV, and $20$ keV,
respectively. The times are listed above each panel: 
$t=4.1$, $4.3$, $4.6$, and $4.85$ Gyr correspond to 
the first maximum of $\beta_{\rm spec}$, the minimum of $\beta_{\rm spec}$, 
the second maximum of $\beta_{\rm spec}$, and the minimum of $T_{\rm e,ew}$,
respectively.
Figure \ref{fig:vmapa} shows the evolution of the gas velocity field,
seen from the same direction at the same times as in figure 
\ref{fig:tbarmapa}. The longest vector corresponds to the maximum velocity 
listed below each panel.

When two subclusters just contact each other at $t=3.5$ Gyr 
(upper left panel), $\bar{T}$ rises slightly at the interface 
between the two subclusters due 
to the adiabatic compression but we can clearly distinguish two 
individual clusters through the X-ray image. Then the two approach 
each other and we see double peaks in the X-ray emissivity profile of 
'one cluster' at $t=4.1$ Gyr (upper middle).
Just between the peaks $\bar{T}$ rises up to $\sim 20$ keV due to a
shock. The shock front is perpendicular to the collision axis. 
However, since $\bar{T}$ still remains nearly the initial value
($\sim 5$ keV) 
around the peaks of the X-ray image, EWMT does not increase
so much that $\beta_{\rm spec}$ becomes rather large (figure \ref{fig:timea}). 
At the most contracting phase ($t=4.3$ Gyr: upper right), 
two peaks merge to one peak in the X-ray image and the X-ray image 
elongates to the directions perpendicular 
to the collision axis. Since the high temperature ($\sim 20$ keV) region is
located around the X-ray peak, EWMT becomes maximum (figure \ref{fig:timea}).
Then the cluster expands and two shocks propagate in the opposite directions
along the collision axis ($t=4.6$ Gyr: lower left). 
We can clearly see two 'lens-shaped' high
temperature regions associated to the shocks in the $\bar{T}$ structure.
We can also see the cooler region ($\sim 5$ keV) spread in the directions
perpendicular to the collision axis. These characteristic temperature 
structures reflect the fact that the gas outflow occurs 
not only along the collision axis 
but also in its perpendicular directions (see figure \ref{fig:vmapa}).
In these directions the gas is effectively cooled by the adiabatic expansion
and emits more X-ray than the hottest regions. Thus, EWMT decreases, 
though high temperature regions still exist. At the most expanding phase
($t=4.85$ Gyr: lower middle), 
the X-ray image elongates along the collision axis. 
The shocks reach in very low density region. We see an 'X-shaped' region
with slightly high temperature (indicated by light blue color). 
This structure is also due to adiabatic expansion
associated to the gas outflows. Then the system contracts again and settles
down to the spherical structure ($t=7$ Gyr: lower right). 
Owing to the subsonic accretion of the gas along the collision axis, 
slightly high temperature regions are created
on the both sides of the collision axis (see figure \ref{fig:vmapa}).

Note that there can be complex temperature structures in ICM when X-ray 
images do not have definite substructures. It is certain that image elongation
is occurred in these case, but this is severely affected by the viewing 
angles. Thus, temperature map is more suitable for the diagnostic of
merging clusters. 

Next, let us see the $T_{\rm e}$ map and discuss the difference between 
$\bar{T}$ and $T_{\rm e}$. Figure \ref{fig:temapa} shows the same as figure
\ref{fig:tbarmapa}, but for $T_{\rm e}$. Although in the contracting phase 
($t \le 4.3$ Gyr) $T_{\rm e}$ maps are similar to the $\bar{T}$ maps,
in the expanding phase ($t>4.3$ Gyr) both maps are rather different.
To see the difference between $\bar{T}$ and $T_{\rm e}$ in detail, we make 
the snapshots of the distribution of 
$\tilde{T}_{\rm e} \equiv T_{\rm e}/\bar{T}$.
Figure \ref{fig:techa} shows the $\tilde{T}_{\rm e}$ maps of Run A.
The red, yellow, green, and blue colors correspond to 
$\tilde{T}_{\rm e} \sim 0.1$, $0.3$, $0.5$, and $0.7$, respectively. 

In the contracting phase ($t \le 4.3$ Gyr), the difference between
$\bar{T}$ and $T_{\rm e}$ is rather small. At $t=4.1$ Gyr, $T_{\rm e}$ 
is slightly lower than $\bar{T}$ at the shock. In overall there are no 
significant two-temperature regions. Then, at $t=4.3$ Gyr, 
split two-temperature regions appear 
due to the propagation of shocks. Since $t_{\rm eq}$ is longer 
in the outer parts where density is low, two-temperature regions are 
more spread there than in the central parts. 
In this phase, however, two-temperature nature 
is less important than in the later phase because the the shocks is 
nearly standing shock and $t_{\rm eq}$ is shorter in the central 
parts.

On the other hand, in the expansion phase ($4.3$ Gyr $\le t \le 4.85$ Gyr), 
the distributions of $\bar{T}$ and $T_e$ are clearly qualitatively 
different. From figure \ref{fig:techa} we find two common features
both at $t=4.6$ Gyr and at $t=4.85$ Gyr. One is that $T_e$ is 
significantly lower ($\sim 50$ \%) than $\bar{T}$ in the post-shock high
$\bar{T}$ regions. These regions are spread on $\sim 0.5$ Mpc scale
behind the shocks. The other is that the difference between $\bar{T}$ and
$T_{\rm e}$ is larger in the directions tilted by the angles of 
$\pm 45^{\circ}$ with respect to the collision axis. 
This is also due to the gas outflows
both along the collision axis and in its perpendicular directions.
The gas originally located in the central part, whose $t_{\rm eq}$ 
is relatively short, moves outward in these directions. 
Note that $t_{\rm eq}$ is hardly changed by the adiabatic compression
and expansion when $\gamma=5/3$.
Thus, the difference between $\bar{T}$ and $T_{\rm e}$ is less 
in these directions. As a result, we see 
four peaks in the $T_{\rm e}$ map at $t=4.6$ Gyr and the 'X-shaped' high 
temperature region becomes less contrasted in the $T_{\rm e}$ map at 
$t=4.85$ Gyr. 
Note that high temperature regions around $\sim 20$ keV cannot be seen 
in the $T_{\rm e}$ maps of the expanding phase. Therefore, it is probable 
that we underestimate the collision velocity by $\sim 30 \%$ 
if we estimate it from the hottest regions in $T_{\rm e}$ map 
in the expanding phase. Two-temperature nature still remains at $t=7$
Gyr in the outer parts as the case of spherical models 
(Fox \& Loeb 1997; Takizawa 1998).

\subsection{Time evolution of temperature distribution of Run B}

First, we describe the dynamical evolution of Run B through
the $\bar{T}$ map and the gas velocity field.
Figure \ref{fig:tbarmapb} shows the same as figure \ref{fig:tbarmapa},
but for Run B. Temperature color scale is adjusted for Run B. 
The blue, green, yellow, and red colors correspond to 
$k_{\rm B} T \sim 4$ keV, $7$ keV, $10$ keV, and $13$ keV, respectively. 
The times are listed above each panel: 
$t=3.9$, $4.1$, $4.3$, and $4.75$ Gyr correspond to the first maximum of 
$\beta_{\rm spec}$, the minimum of $\beta_{\rm spec}$, 
the second maximum of $\beta_{\rm spec}$, and the minimum of $T_{\rm e,ew}$,
respectively. Figure \ref{fig:vmapb} shows the same as figure \ref{fig:vmapa},
but for Run B. 

When the two subclusters approach each other, 
the bow shock with an arc shape 
is formed just between the two. Thus, the arc-shaped hot region is
seen between the two peaks of the X-ray image at $t=3.9$ Gyr (upper middle). 
This region clearly separates the component of larger subcluster 
($\sim 5$ keV) from the component of smaller subcluster ($\sim 2.5$ keV).
Then, the two peaks merge to one triangle image of X-ray 
($t=4.1$ Gyr: upper right). 
The hottest temperature region ($\sim 13$ keV) is located 
in the peak of the X-ray image and the hot region associated to the 
bow shock is seen elongated slantingly backward with respect to the
motion of the smaller subclusters. Then the gas expands and the
two shocks propagate outward along the collision axis. The one 
in front of the motion of the smaller subcluster 
(for the left of the figure) is arc-shaped and the other 
is rather flat ($t=4.3$ Gyr: lower left). 
At the most expanding phase ($t=4.75$ Gyr: lower middle) the X-ray 
image elongates along the collision axis and the shocks reach very low 
density regions. The elongation of the image is more significant
for the left of the figure.
Then the system contracts again and settles down to the
spherical structure ($t=7$ Gyr: lower right).
Accretion flows are seen only from the front side with respect to the
motion of the smaller subcluster.

Next, we describe two-temperature nature of Run B through
the $T_{\rm e}$ map and the $\tilde{T}_{\rm e}$ map.
Figure \ref{fig:temapb} shows the same as figure \ref{fig:tbarmapb}, 
but for $T_{\rm e}$ and figure \ref{fig:techb} shows the same as 
figure \ref{fig:techa}, but for Run B.
As the case of Run A, in the contracting phase ($t \le 4.1$ Gyr)
the $T_{\rm e}$ distribution is similar to the $\bar{T}$ one 
though $T_{\rm e}$ is slightly lower than $\bar{T}$ around the shock.
On the other hand, in the expanding phase ($t > 4.1$ Gyr) 
substantial deviation emerges especially around the hot regions associated
to the shocks. At $t=4.3$ Gyr in the $\bar{T}$ map both two hot regions
have almost the same temperature ($\sim 13$ keV), whereas in the $T_{\rm e}$
map, the former (with respect to the motion of the smaller subcluster) 
hot region becomes less than $\sim 10$ keV. On the other hand, in the backward 
hot region, $\sim 13$ keV component still remains. In this region,
$t_{\rm eq}$ is rather short because the gas there originates from the smaller 
subcluster (see the scaling relation (\ref{eq:screlteq})). 
At $t=4.75$ Gyr, as the case of Run A, 
$\tilde{T}_{\rm e}$ is low in the directions tilted by the angles of 
$\pm 45^{\circ}$ with respect to the collision axis, although this 
tendency is less significant than in Run A.

\section{SUMMARY AND DISCUSSION}\label{s:sd}

We investigate evolution of ICM during mergers considering the 
relaxation process between the ions and electrons. From the simple
analytical estimation, we find that in the CDM universe the equilibrium 
timescale between ions and electrons are longer in more massive CG 
even if the gas fraction is dependent on CG mass in such a way that
the X-ray luminosity is proportional to the cube of the temperature.
We estimate the collision velocity and show that supersonic collisions are 
quite natural in CDM universe in major mergers and that electron 
temperature can be less than a half of the plasma mean temperature
in the post-shock regions. Temperature estimated from X-ray observations 
can be significantly lower than ion temperature.

We carry out the numerical simulations of the mergers by N-body
and hydrodynamical simulations, incorporating the relaxation process 
between the ions and electrons. To solve the evolution of the normalized
electron temperature, we adopt the results of Fox \& Loeb (1997).
The difference in emissivity-weighted mean temperatures of ions
and electrons is practically negligible during the merger; the discrepancy
is $\sim 1 \%$ even at maximum. On the other hand, the spatial 
distribution of the electron temperature is significantly different
from that of the plasma mean one especially in low-density regions
in the expanding phase. 
In this phase, electron temperature is at most $\sim 50\%$ lower than 
the plasma mean one in the post-shock hot regions. In addition,
the difference between them is more enhanced in the directions tilted 
by the angles of $\pm 45^{\circ}$ with respect to the collision axis.
When the two subclusters have different masses, the hot region
located to the former position (with respect to the direction of the 
motion of the smaller subcluster) has lower electron temperature than that
located backward in the expanding phase.

Recently, Markevitch et al. (1998) estimate the subcluster
collision velocity of Cygnus A using the electron temperature map obtained 
by ASCA. From this temperature map and the ROSAT PSPC image, it is likely
that Cygnus A is in just contracting phase such as Run A at $t=4.1$ Gyr.
Thus, two-temperature nature is almost negligible in this case.
However, collision velocities are likely underestimated by $\sim 30 \%$
if we use the electron temperature of the hottest regions of merging 
clusters in the expanding phase such as SC1329-313 (Hanami et al. 1998).

Although we consider only head-on collisions for simplicity
in the present calculations , off-center collisions should be 
investigated to model
more realistic situations. Ricker (1998) find that that spiral bow shocks 
occur in off-center mergers, which probably produce rather complex electron
temperature distribution. We will investigate this issue as a future work.

We consider only the classical Coulomb coupling as the relaxation process
between ions and electrons. It is possible, however, in ICM more efficient
relaxation processes could be effective (McKee \& Cowie 1977; 
Pistinner, Levinson, \& Eichler 1996). In this case 
the equilibrium timescale could be shorter than the value given by equation
(\ref{eq:teq}). Therefore, the temperature difference between ions and
electrons can be less than our results. If magnetic field exits in ICM, 
it is possible that electrons are also significantly heated by shocks 
associated with MHD instabilities. 
Also in this case the temperature difference
between ions and electrons could be less.

We neglect the heating process from the galaxies to ICM. 
If substantial amount of thermalized hot gas is injected into ICM 
from the galaxies, temperature difference could be less. 
It is possible that mergers of CG activate the star formation 
in the member galaxies through the
galaxy-galaxy interaction, galaxy-ICM interaction, etc
(Caldwell et al. 1993) . 
In this case, this effect cannot be negligible but the effect on ICM
temperature distribution is rather sensitive on the detailed modeling 
of the star formation activity (Fujita 1998).

We adopt the SPH method for the hydrodynamical part in the present 
simulations for simplicity to treat the coupling between ions and
electrons. However, resolution of the shock in SPH is not better than 
that in high resolution finite-difference codes such as TVD, PPM, CIP, etc,
because artificial viscosity is used to treat the shocks in SPH. 
Furthermore, bulk viscosity is included in our code and possibly effects
electron temperature slightly. Although we believe
that our simulations can follow the qualitative behavior, it is 
useful to simulate two-temperature ICM during mergers 
with finite-difference codes.

\acknowledgements

The author would like to thank C. Ishizaka, H. Hanami, and S. Mineshige
for helpful comments. This work is supported in part by a Grant-in-Aid from 
the Ministry of Education, 
Science, Sports and Culture of Japan (6179) and
Research Fellowships of the Japan Society for the 
Promotion of Science for Young Scientists.

\appendix
\section{DERIVATION OF THE EQUATION FOR THE NORMALIZED ELECTRON TEMPERATURE}
In the SPH calculations shocks are treated through artificial viscosity.
We assume that artificial viscous heating is effective only for ions
and that only Coulomb coupling is considered in the relaxation process.
Then the Laglangean time evolution of the electron temperature $T_{\rm e}$
and the mean temperature $\bar{T} = (n_{\rm e} T_{\rm e} + 
n_{\rm i} T_{\rm i})/(n_{\rm e} + n_{\rm i})$ is
\begin{eqnarray}
	\frac{d T_{\rm e}}{dt} &=& (\gamma-1)\frac{T_{\rm e}}{n}\frac{dn}{dt}
              + \frac{T_{\rm i} - T_{\rm e}}{t_{\rm eq}}, \\
        \frac{d \bar{T}}{dt} &=& (\gamma-1)\frac{T}{n}\frac{dn}{dt} 
              + \frac{2}{3} \frac{\mu m_{\rm p}}{k_{\rm B}} Q_{\rm vis},
\end{eqnarray}
where $T_{\rm i}$ is the ion temperature, $n$ is the gas density, and
$\gamma=5/3$ is the ratio of specific heats and $Q_{\rm vis}$ is 
the artificial viscous heating per unit mass.
We use the artificial viscosity in the form described in 
Monaghan (1992) \S4.1, which is the most commonly used one.
Then, the explicit expression of $Q_{\rm vis}$ for the $i$-th SPH 
particle is
\begin{eqnarray}
	Q_{{\rm vis,}i} = \frac{1}{2} \sum_j m_j \Pi_{ij}
         ( \mbox{\boldmath $v$}_i-\mbox{\boldmath $v$}_j ) \cdot
         \nabla_i W_{ij},
\end{eqnarray}
where $\Pi_{ij}$ is given by
\begin{eqnarray}
	 \Pi_{ij} = \left\{
                      \begin{array}{@{\,}ll}
	\frac{-\alpha \bar{c}_{ij}\mu_{ij} + \beta \mu_{ij}^2}
             { \bar{\rho}_{ij} }, &
        {\rm for \ \ } (\mbox{\boldmath $v$}_i-\mbox{\boldmath $v$}_j) \cdot
        (\mbox{\boldmath $r$}_i-\mbox{\boldmath $r$}_j) < 0, \\
        0, &
        {\rm for \ \ } (\mbox{\boldmath $v$}_i-\mbox{\boldmath $v$}_j) \cdot
        (\mbox{\boldmath $r$}_i-\mbox{\boldmath $r$}_j) >0,
                      \end{array}
                      \right. 
\end{eqnarray}
and
\begin{eqnarray}
	\mu_{ij} = \frac{h_{ij}
        (\mbox{\boldmath $v$}_i-\mbox{\boldmath $v$}_j) \cdot
        (\mbox{\boldmath $r$}_i-\mbox{\boldmath $r$}_j) }
        {(\mbox{\boldmath $r$}_i-\mbox{\boldmath $r$}_j)^2
         + \eta^2}.
\end{eqnarray}

In the above expressions, $m_j$, $\mbox{\boldmath $v$}_j$, and
$\mbox{\boldmath $r$}_j$ are the mass, velocity, and position of the 
$j$-th SPH particle, respectively.  $c_{ij}$, $\rho_{ij}$ and $h_{ij}$
are the average of sound speed, density, and smoothing length of
particle $i$ and $j$, respectively. We set $\alpha=1$, $\beta=2$, and
$\eta^2 = 0.01$, which are typical values in SPH calculations.
$\nabla_i W_{ij}$ is the gradient of the $j$-th particle's 
kernel function at $\mbox{\boldmath $r$}_i$.

Introducing the temperature normalized by $\bar{T}$, i.e., 
$\tilde{T_{\rm e}} \equiv (T_{\rm e}/\bar{T})$ and 
$\tilde{T_{\rm i}} \equiv (T_{\rm i}/\bar{T})$, we find
\begin{eqnarray}
	\frac{d \tilde{T_{\rm e}}}{dt} = 
        \frac{\tilde{T_{\rm i}}-\tilde{T_{\rm e}}}{t_{\rm eq}}
        -\tilde{T_{\rm e}}\frac{Q_{\rm vis}}{u},
\end{eqnarray}
where $u \equiv 3 k_{\rm B} \bar{T} / (2 \mu m_{\rm p})$ is the thermal
energy per unit mass.

\clearpage
 \begin{table}
  \begin{center}
   \begin{tabular}{ccc} 
    \hline \hline
                             &        Run A          &       Run B        \\ 
    \hline
$M_1/M_2 (10^{15} M_{\odot}$)&      0.5/0.5          &      0.5/0.125     \\
$r_{{\rm c},1}/r_{{\rm c},2}$  (Mpc)&  0.2/0.2       &      0.2/0.1       \\
$k_{\rm B}T_1/k_{\rm B}T_2$ (keV)&  4.78/4.78        &      4.78/2.39     \\
$r_{\rm sep}$ (Mpc)          &      4.0              &       3.3          \\
$f_{\rm g}$                  &      0.1              &       0.1          \\
$\epsilon$ (Mpc)             &      0.02             &       0.01         \\
$N_1/N_2$ (SPH)              &     5000/5000         &     5000/1250      \\
$N_1/N_2$ (DM)               &     5000/5000         &     5000/1250      \\
    \hline
   \end{tabular}
  \end{center}
  \caption[Model parameters]{Model parameters}
  \label{tab:para}
 \end{table}%

\newpage

\figcaption{The square of Mach numbers of two virialized clusters 
as functions of $\alpha$ (the ratio of the two
cluster masses) during the merger for the case with $P(k) \propto k^{-2}$. 
The solid and dotted lines show the values for the larger
and the smaller cluster, respectively.
\label{fig:mach}}

\figcaption{The time evolution of various physical 
quantities representing the features of the entire system for Run A:
 from the top, (a) the total luminosity ($L$), 
(b) the emissivity-weighted mean electron temperature ($T_{\rm e,ew}$), 
(c) the square root of the velocity 
dispersion parallel to the collision axis ($\sigma_{//}$), and (d) 
the specific energy ratio of DM and ICM ($\beta_{\rm spec}$). 
\label{fig:timea}}

\figcaption{The snapshots of X-ray surface brightness (contours)
overlaid with emissivity-weighted $\bar{T}$ (colors) distribution seen 
from the direction perpendicular to the collision axis for Run A. 
X-ray surface brightness contours are equally spaced on a logarithmic scale
and separated by a factor of 7.4. The blue, green, yellow, and red
colors correspond to $k_{\rm B} T \sim 5$ keV, $10$ keV, $15$ keV, and 
$20$ keV, respectively. The times are listed above each panel.
\label{fig:tbarmapa}}

\figcaption{The evolution of the gas velocity field 
for Run A seen from the same direction at the same times as in figure 
\ref{fig:tbarmapa}. The longest vector corresponds to the maximum velocity 
listed below each panel.
\label{fig:vmapa}}

\figcaption{Same as figure \ref{fig:tbarmapa}, but for $T_{\rm e}$ 
instead of $\bar{T}$ for Run A
\label{fig:temapa}}

\figcaption{The snapshots of distribution of normalized electron
temperature, $\tilde{T_{\rm e}} \equiv T_{\rm e}/\bar{T}$ 
for Run A. The red, yellow, green, and blue colors correspond to 
$\tilde{T}_{\rm e} \sim 0.1$, $0.3$, $0.5$, and $0.7$, respectively. 
\label{fig:techa}}

\figcaption{Same as figure \ref{fig:timea}, but for Run B.
\label{fig:timeb}}

\figcaption{Same as figure \ref{fig:tbarmapa}, but for Run B.
Temperature color scale is adjusted for Run B. 
The blue, green, yellow, and red colors correspond to 
$k_{\rm B} T \sim 4$ keV, $7$ keV, $10$ keV, and $13$ keV, respectively. 
\label{fig:tbarmapb}}

\figcaption{Same as figure \ref{fig:vmapa}, but for Run B.
\label{fig:vmapb}}

\figcaption{Same as figure \ref{fig:temapa}, but for Run B.
\label{fig:temapb}}

\figcaption{Same as figure \ref{fig:techa}, but for Run B.
\label{fig:techb}}

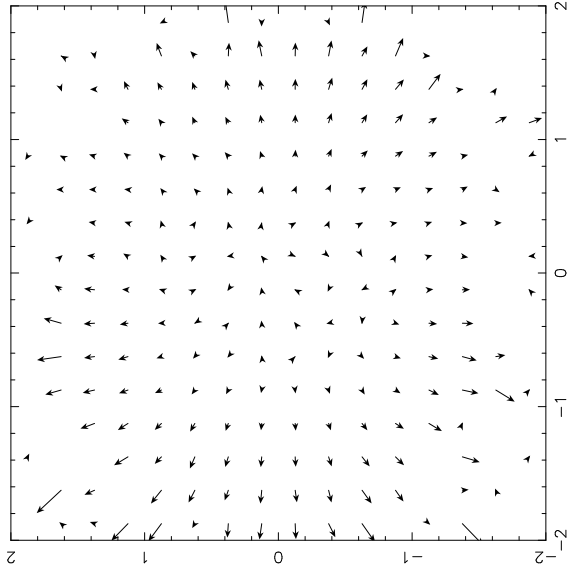
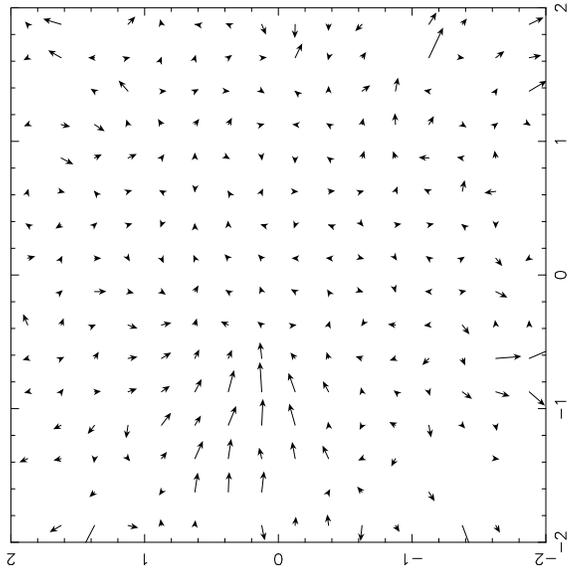
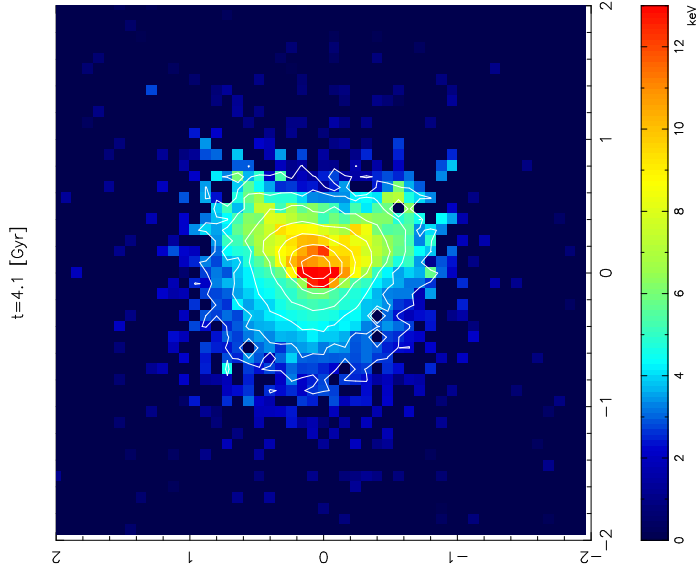
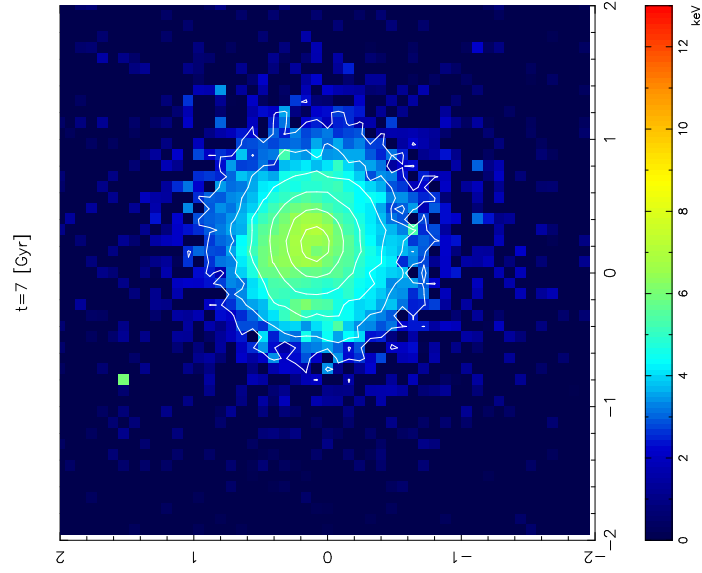
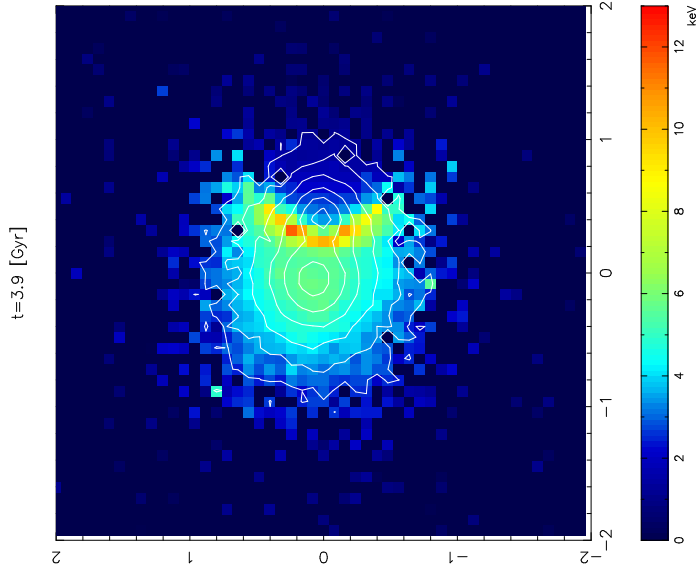
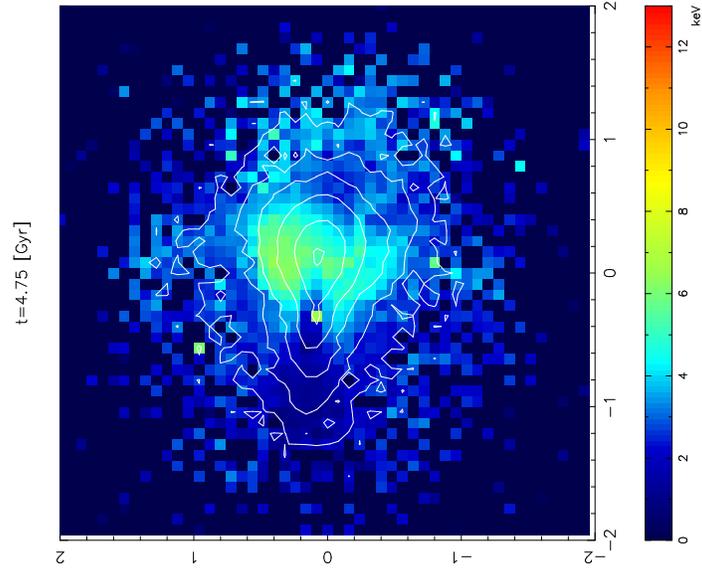
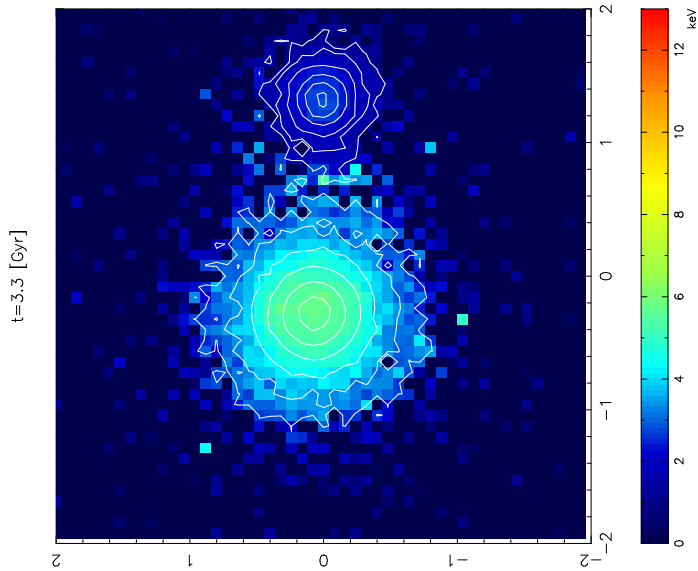
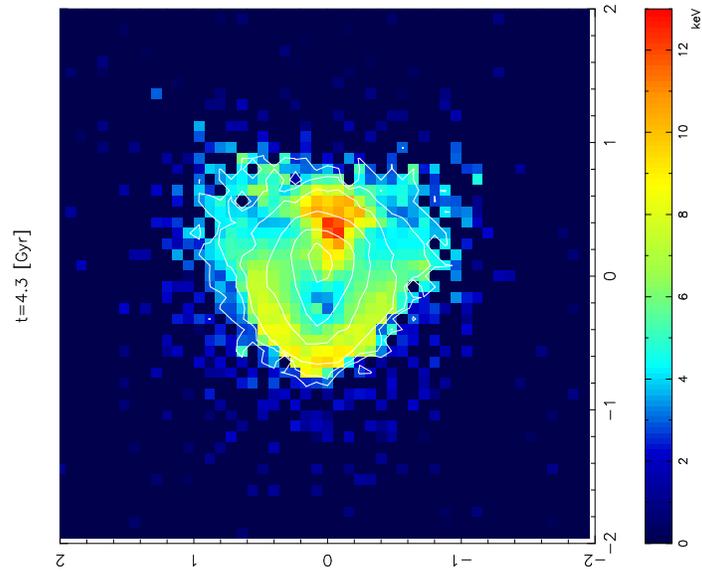
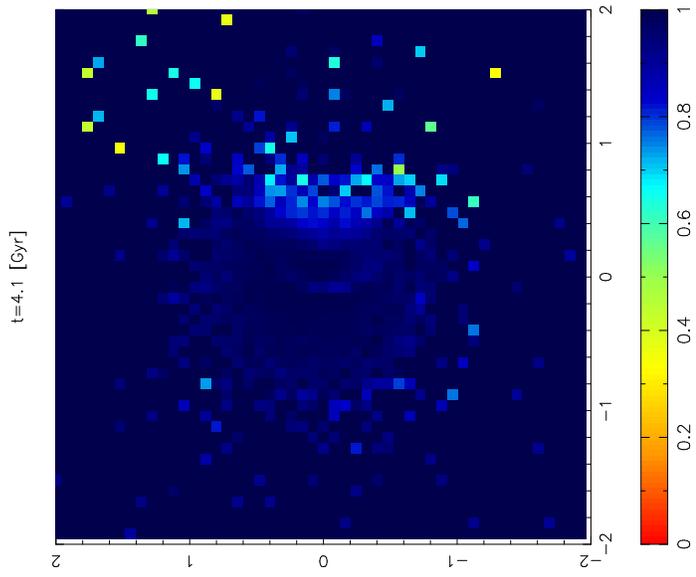
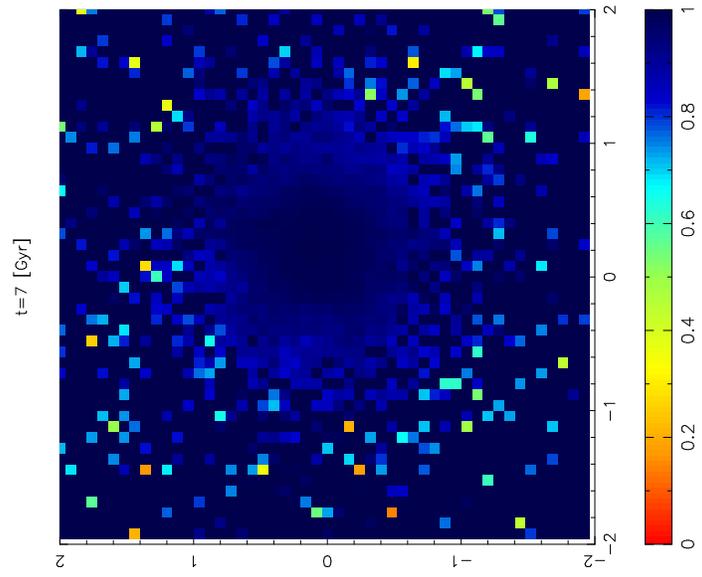
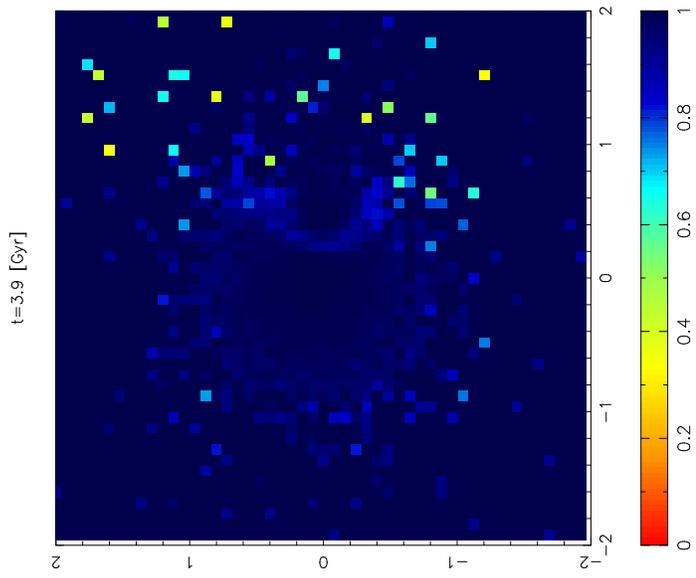
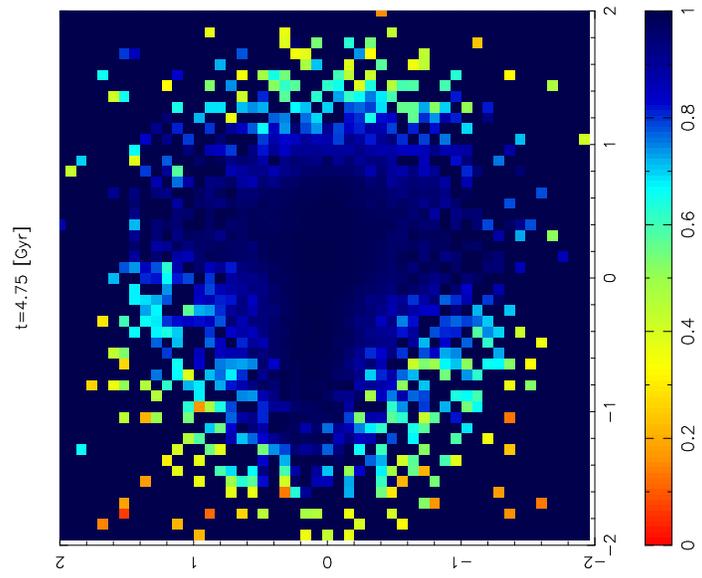
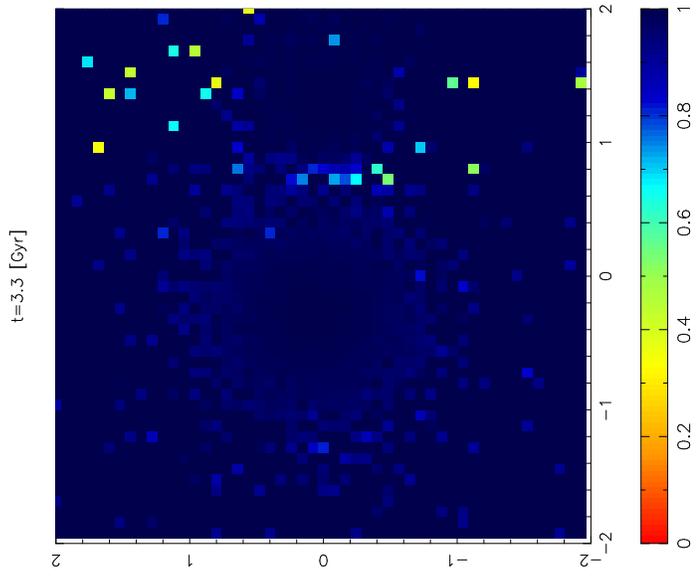
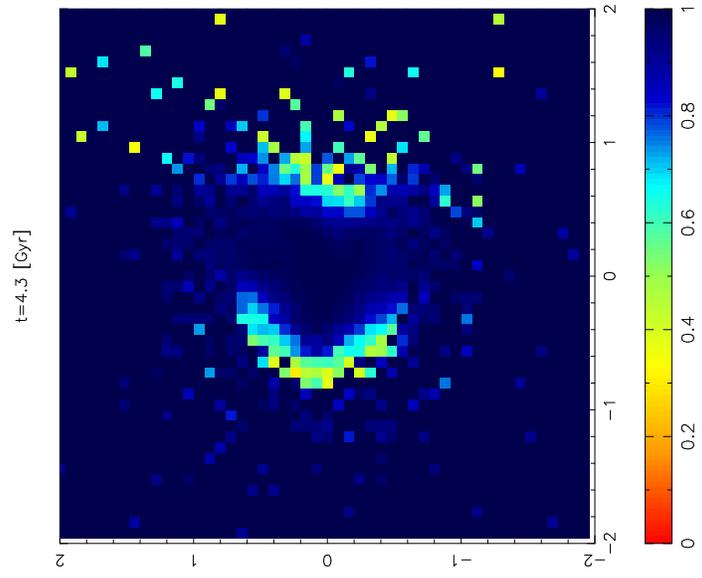
\end{document}